\documentclass[default,iicol]{sn-jnl}

\usepackage{accents}
\usepackage{siunitx}

\usepackage{subcaption}
\usepackage{multicol}

\jyear{2021}%

\theoremstyle{thmstyleone}%
\theoremstyle{thmstyletwo}%

\theoremstyle{thmstylethree}%

\begin{document}

\title[Article Title]{Two-Photon Optical Ramsey-Doppler Spectroscopy of Positronium and Muonium}


\author[1]{\fnm{Evans} \sur{Javary} }\email{ejavary@ethz.ch}
\author[1]{\fnm{Edward} \sur{Thorpe-Woods} }\email{ethorpe@ethz.ch}
\author[1]{\fnm{Irene} \sur{Cortinovis} }\email{coirene@ethz.ch}
\author[1]{\fnm{Marcus} \sur{Mähring} }\email{mmaehring@ethz.ch}
\author[1]{\fnm{Lucas} \sur{de Sousa Borges} }\email{lucasde@ethz.ch}
\author*[1]{\fnm{Paolo} \sur{Crivelli}} \email{paolo.crivelli@cern.ch}

\affil[1]{\orgdiv{Institute for Particle Physics and Astrophysics}, \orgname{\textsc{eth}}, \orgaddress{\city{Zurich}, \postcode{8093}, \country{Switzerland}}}

\abstract{

Positronium and muonium, as purely leptonic atoms without internal structure, provide ideal systems for high-precision tests of quantum electrodynamics (QED) and measurements of fundamental constants. However, the high velocities of these lightweight atoms complicate precision spectroscopy, particularly in the 1S-2S transition, due to transit time broadening and second-order Doppler shifts. To overcome these challenges, we propose a novel method combining two-photon Ramsey spectroscopy with a technique to correct the second-order Doppler shifts on an atom-by-atom basis. Additionally, this approach suppresses systematic effects of the AC Stark shift to a negligible level compared to the target precision. Simulations predict that for both positronium and muonium, this method could improve the measurement precision of the 1S-2S transition by more than two orders of magnitude compared to the current state of the art. This approach opens up new avenues for rigorous bound-state QED tests and searches for physics beyond the Standard Model.
}

\maketitle

\section{Introduction}\label{sec1}

Positronium (Ps), the bound state of an electron ($e^-$) and a positron ($e^+$), and Muonium (M), the bound state of an antimuon  ($\mu^+$) and an electron, are purely leptonic atoms. As the constituents lack internal structure, they are free from finite-size effects. These atoms, due to their simplicity, provide excellent systems for precision tests of bound state QED \cite{2005_Karshenboim, adkins2022precision} and the determination of fundamental constants such as the muon mass. They are also unique probes to search for potential new physics beyond the Standard Model. Those include searches for new exotic forces \cite{Cong:2024qly}, dark sectors \cite{Frugiuele_2019,Stadnik:2022gaa}, Lorentz and CPT violation \cite{2014_Vargas} or the effect of gravity on antimatter \cite{Karshenboim:2008zj}. 
Although leptonic systems are well understood theoretically, precise spectroscopy measurements remain a highly challenging task, primarily due to their very short lifetimes: \SI{142}{ns} for Ps originating from the e$^+$e$^-$ annihilation in the triplet spin state and \SI{2.2}{\micro s} for M, limited by the muon lifetime. Furthermore, the significantly lighter mass of Ps and M 
($2m_e$ and $\sim208m_e$)
compared to regular atoms means that they possess high velocities even at low kinetic energies. 

Nevertheless, precise spectroscopic measurement of the hyperfine structure \cite{liu1999high,strasser2019new,2020_MUSEUM}, the fine structure \cite{PhysRevLett.131.043001} and Lamb shift \cite{Mu-MASS:2021uou} have been conducted.  The most precise measurement of the positronium 1S-2S transition frequency is 1233607216.4(3.2) MHz \cite{Fee1993}, which is in agreement at the level of 1.8$\sigma$ with the QED prediction of \SI{1233607222.12 \pm  0.58}{MHz} \cite{adkins2022precision}.
Similarly, the current best measurement of the muonium 1S-2S transition, 2455528941.0(9.8)MHz \cite{2000_Meyer}, aligns well with the QED prediction of 2455528935.2(1.4)MHz \cite{cortinovis2023update}.

In this work we outline a technique with the potential to improve by orders of magnitude the 1S-2S transition in positronium and muonium using a variation of Ramsey spectroscopy \cite{Ramsey_OG}, which we refer to as Ramsey-Doppler spectroscopy. This novel technique combines Ramsey spectroscopy methods with a detection scheme which accounts for the second-order Doppler effect. Even though two-photon spectroscopy using a Fabry-Perot cavity is free from the first order Doppler effect, the relativistic second-order Doppler effect results in a shift which is $\delta\nu_{2DS}=-(v^2/2c^2)\nu_0$, where $v$ is the atom's velocity, $c$ the speed of light and $\nu_0$ the transition frequency.
The Ramsey-Doppler method can be used to overcome the limitations for precision spectroscopy of exotic atoms, reducing the measured linewidth caused by finite laser transit time, systematic uncertainties stemming from the AC Stark effect, and correcting the second-order Doppler effect on an atom-by-atom basis. 
Monte Carlo simulations of the scheme proposed here predict a significantly improved experimental determination of the 1S-2S transition in positronium and muonium, with a precision that would exceed current theoretical predictions.

\section{Two-Photon Optical Ramsey-Doppler Scheme}
Ramsey spectroscopy has proven to be a valuable tool in the RF regime \cite{Ramsey_OG}, serving as the operation principle for many atomic clocks. In standard Rabi spectroscopy, atoms are exposed to resonant electromagnetic radiation within a single `interaction region', during which the electronic transition of interest can be excited. The fundamental principle of Ramsey spectroscopy involves replacing a single interaction region with two spatially separated interaction regions that oscillate phase coherently. When atoms pass through the first region, they are placed into a coherent superposition of the ground and excited states. The atoms then continue their trajectory between the two interaction regions, during which their quantum states evolve freely. Upon traversing the second interaction region, they interact with the field again and are either excited or de-excited, depending on the relative phase difference between the atoms and the field. The result is a spectrum that has distinctive Ramsey fringes, the linewidth of the central fringe is inversely proportional to the free evolution time between the two interaction zones. As the duration of free evolution can be more easily extended than the interaction time, the Ramsey technique offers a straightforward method to reduce the measured line widths.

\begin{figure*}[htp]
    \centering
    \includegraphics[width=0.9\linewidth]{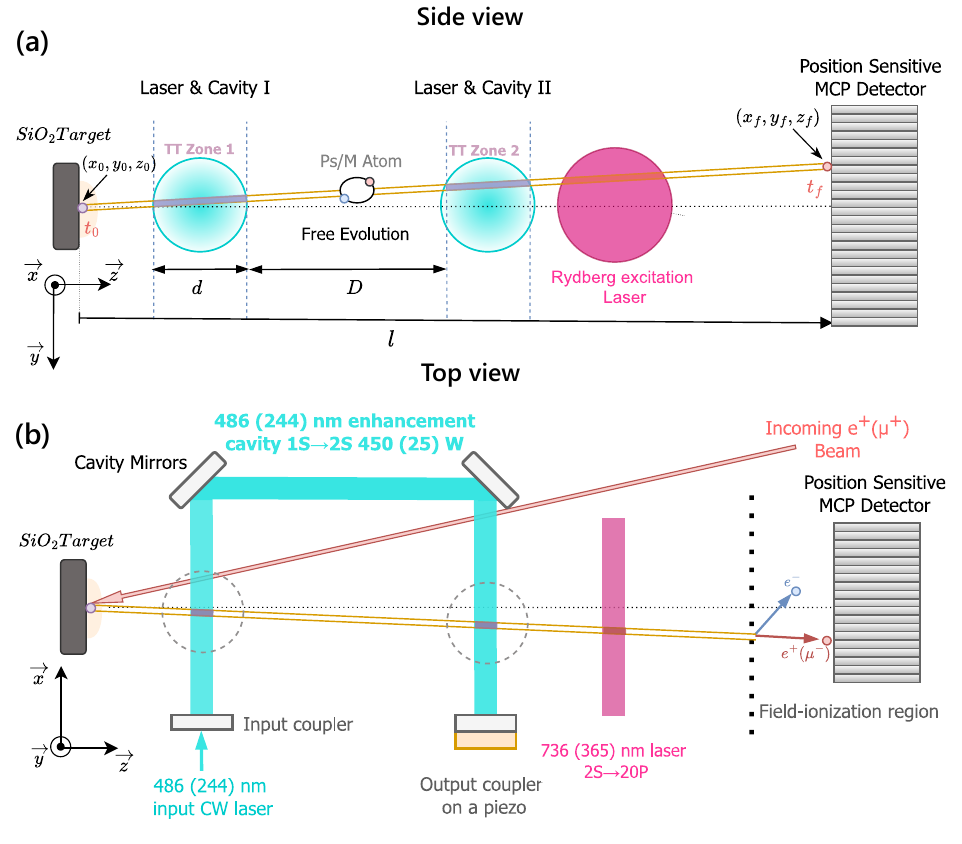}
    \caption{ Side (a) and top views (b) of the schematic of the 1S-2S Ramsey-Doppler spectroscopy setup for positronium and muonium atoms. Ps/M atoms are formed when a $e^+$/$\mu^+$ beam impinges on a $\text{SiO}_2$ target. Atoms diffuse out of the target and traverse the first laser interaction region, which places them in a superposition of 1S and 2S states. The atom wavefunction is then allowed to freely evolve, and the phase accumulates between the atom and the field. Atoms are subsequently either excited or de-excited by the second interaction region. Atoms in the 2S state are excited to the 20P Rydberg state by a second pulsed laser, and subsequently field-ionised and detected by a position-sensitive microchannel plate (MCP) \cite{Heiss:2024ofn}. The position-sensitive detection allows the atom's trajectory to be reconstructed, estimating the velocity to correct for the second-order Doppler effect.}
    \label{fig:setup}
\end{figure*}

To apply the Ramsey technique at optical wavelengths, a major challenge originates from keeping a well defined phase relation between both excitation regions. For a running wave, the phase varies in space on the scale of the optical wavelength which is significantly smaller than the interaction region. This causes the fringe pattern to wash out. In contrast, for two-photon spectroscopy in a standing wave as is proposed here, the phase is space-independent. Optical two-photon Ramsey excitation with two spatially separated interaction regions was proposed and realised experimentally for hydrogen \cite{Huber1998TWOPHOTONOR}. More recently, a novel scheme employing Ramsey-comb spectroscopy was demonstrated, opening the possibility for high-precision spectroscopy of molecular hydrogen and helium ions \cite{PhysRevLett.117.173201}. This technique addresses the experimental challenges posed by the requirement for deep ultraviolet or even shorter wavelengths.
 
In the following subsections, we outline experimental schemes for conducting two-photon Ramsey spectroscopy of the 1S-2S transition in positronium and muonium. Optical Ramsey spectroscopy has not yet been conducted in leptonic atoms, despite its potential benefits; although it presents some experimental challenges, it also provides unique advantages. Transitioning from standard Rabi spectroscopy to the Ramsey scheme reduces the experimental signal rate; however, it improves spectral resolution by producing interference fringes. This significantly reduces transit time broadening, thereby improving the precision of determining the central frequency. Another benefit of the Ramsey spectroscopy method is that it suppresses the AC Stark shift on the transition frequency \cite{borde1983density_ac_suppr} as detailed in section \ref{sec:syst}. 

The experimental setup for conducting Ramsey spectroscopy for positronium and muonium is broadly the same and is illustrated on Figure \ref{fig:setup}.  The specifics of the setups for Ps and M are detailed in sections \ref{sec: pos} and \ref{sec:mu} respectively. To produce Ps/M atoms, a beam of positrons/muons is directed on a porous silica target, where the leptonic atoms are formed \cite{Crivelli2010,2012-Mesoporous}. Atoms diffuse out of the silica target and traverse two identical, phase-coherent interaction regions. The two interaction regions are produced by folding a laser enhancement cavity using highly reflective mirrors. After passing through the two interaction regions, the atoms will cross a third laser beam, which excites atoms in the 2S state to the 20P Rydberg state. The atoms then continue their trajectory to the detection region, where those which have been excited to Ryberg states are field-ionised and detected using a position-sensitive MCP \cite{Heiss:2024ofn}. The position-sensitive detection records the coordinates $(x_f, y_f)$  and time $(T_f)$ at which the ion from field-ionisation is detected. This information can be used to reconstruct the velocity $(v_x, v_y, v_z)$ of the atoms, when combined with the formation position $(x_0, y_0, z_0)$ 
and the formation time $(T_0)$ (which is determined differently in positronium and muonium as presented in \ref{sec: pos} and \ref{sec:mu} respectively). 

The reconstructed velocity of the atom is used to correct for the second-order Doppler shift on an atom-by-atom basis (see section \ref{sec: Doppler correction}). This method, which we refer to as the Ramsey-Doppler technique, provides a novel approach for tracing out a Ramsey spectrum using the second-order Doppler effect with a fixed laser frequency. The laser frequency experienced by the atom in its frame of reference depends on its velocity in the lab frame, due to the second-order Doppler effect. Therefore, if the second-order Doppler effect can be precisely accounted for, then it can also be used as a mechanism for scanning the laser detuning during spectroscopy. By plotting the detected events vs. the detuning in the positronium frame of reference, it is possible to trace a Ramsey spectrum for the 1S-2S transition of positronium without varying the laser frequency in the lab frame.

\subsection{Scheme for Positronium}\label{sec: pos}

The Ramsey-Doppler scheme for positronium is visualised in Figure \ref{fig:setup}. To study its implementation we use the parameters of the pulsed-mode low-energy positron beamline situated at the Laboratory for Positron and Positronium Physics at ETH Zurich. This beamline was used to develop the detection method based on the 1S-2S and 2S-20P positronium pulsed excitation employing field-ionisation for velocity reconstruction \cite{Heiss:2024ofn}. The system is currently being upgraded with a continuous-wave (CW) laser to conduct standard Rabi spectroscopy. 

Positronium atoms are produced by directing the positrons onto a porous silica target \cite{Crivelli2010}, using the timing of the beam pulse as the reference for the positronium formation time, $T_0$. The Ps atoms will traverse a twice-folded Fabry-Perot optical cavity, which enhances \SI{486}{nm} CW laser light to over \SI{450}{W}. The positronium is subsequently excited from the 2S to the 20P state using a \SI{736}{nm} pulsed laser, which is then field ionised and detected using a position-sensitive MCP.

\subsection{Scheme for Muonium}\label{sec:mu}

The Ramsey-Doppler scheme for muonium is also visualised in Figure \ref{fig:setup}. The experimental setup is planned for implementation within the Mu-MASS experiment  at the Paul Scherrer Institute (PSI) \cite{2018_Crivelli, cortinovis2023update}. Mu-MASS currently employs the low energy muon (LEM) beamline where the incoming $\mu^+$ is tagged by detecting secondary electrons emitted as it passes through a thin carbon foil \cite{janka2024improving}. This allows for the determination of the time, $T_0$, at which muonium is formed in the target. The two interaction regions are produced from a twice-folded Fabry-Perot optical cavity, which enhances \SI{244}{nm} CW laser light to approximately \SI{25}{W}, using techniques developed in  \cite{Burkley:21, Zhadnov:23} while the Rydberg excitation is conducted with a pulsed \SI{365}{nm} laser. 

For the realisation of the proposed experiment with M, higher muon rates than what is currently available are needed. Thus, the proposed method is reliant on the planned High Intensity Muon Beamline (HIMB) at PSI, promising to deliver ${10^{10}}\SI{}{\mu^+/s}$---an improvement of two orders of magnitude over the state of the art \cite{HIMB_Status_2023}---and its combination with MuCool, a novel technique to efficiently slow down and focus muons down to a sub-millimeter beam \cite{antognini2021mucool}. For example, using a SiO$_2$ target of size 10x1 mm\textsuperscript{2}, the predicted rate is $3 \times 10^4 $ muonium atoms per second (see Table \ref{tab:PredMRates}). We will also consider the option of using a superfluid helium (SFHe) target to produce a mono-energetic muonium beam as discussed in section \ref{SFHe}.

\begin{figure*}[h!]
    \centering
    \includegraphics[width=\linewidth]{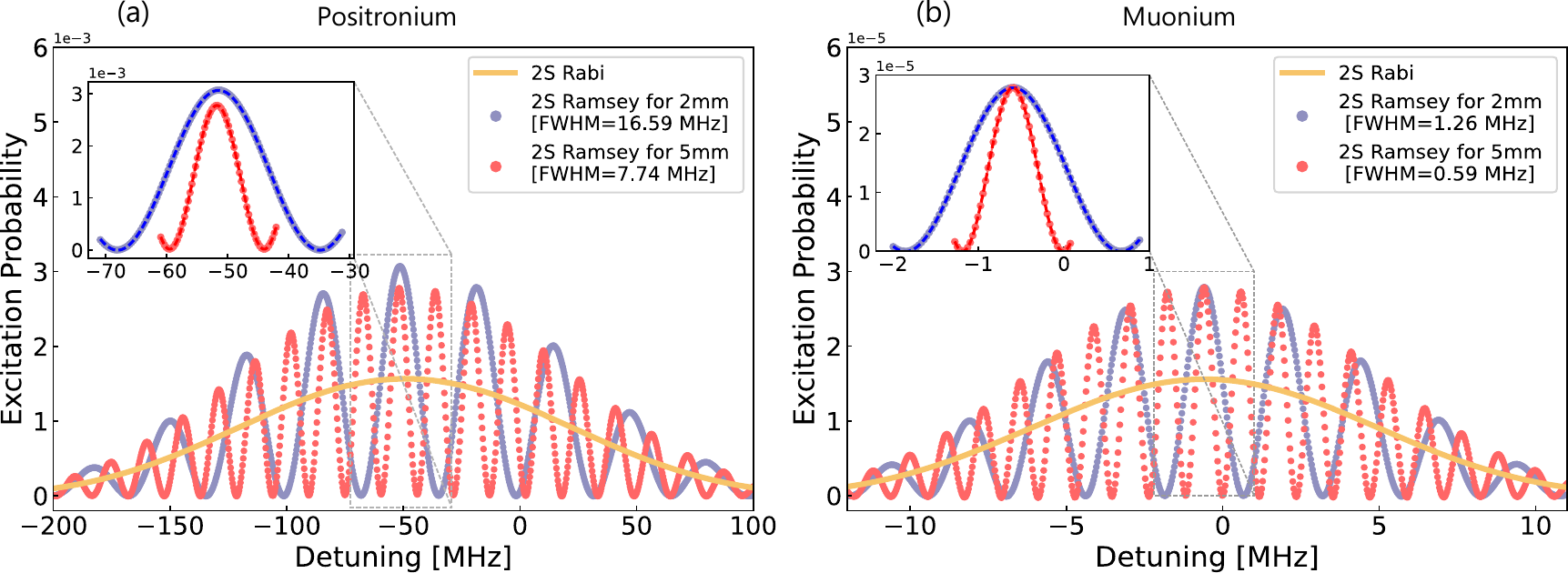}
    \caption{Comparison of the full width at half maximum (FWHM) of the 2S state of positronium and muonium as a function of the detuning in MHz for Rabi and Ramsey configurations. (a) For positronium, the Rabi and Ramsey configurations with free evolution lengths of 2 mm and 5 mm yield FWHM values of 16.59 MHz and 7.74 MHz, respectively, demonstrating the improved resolution of the Ramsey method. (b) For muonium, the FWHM values are 2.16 MHz for 2 mm and 0.59 MHz for 5 mm.}
    \label{fig:Ramsey fringes}
\end{figure*}

\section{Monte Carlo Simulation }\label{MonteCarloSim}
The Ps and M atoms are initialised within the simulation at time $T_0$, with an initial location ($x_0, y_0, z_0$), and velocity components \((v_x, v_y, v_z)\). These parameters are sampled from predefined distributions---cosine for trajectories, Maxwell-Boltzmann for velocities, and Gaussian for spatial and temporal profiles---to simulate realistic experimental conditions. As the atoms are emitted from the source, they propagate through the two interaction zones predominantly along the z-direction. The trajectory of the atoms is simulated on the basis of its initial position and velocity. The optical Bloch equations are numerically integrated at every time step of the simulation to track the evolution of the atom's state. The trajectory of the atom terminates in the plane that contains the MCP, the final probability of excitation is determined from the optical Bloch equation parameters and recorded, along with the final coordinates ($x_f, y_f, z_f$) and time $T_f$ of the trajectory. A Gaussian smear is added to the final spatial and temporal coordinates to account for MCP detector resolution.
 
 The input parameters for the simulation are based on the knowledge from previous experiments at ETH Zurich \cite{Heiss:2021qxv} and PSI \cite{cortinovis2023update}. These inputs include the velocity distribution of the atoms, the parameters of the laser beams, the distances between the lasers, the target and the source, as well as the spatial distribution of the beam spot and trajectories of the atoms.

\begin{figure*}[htp]  
    \centering
    \includegraphics[width=\linewidth]{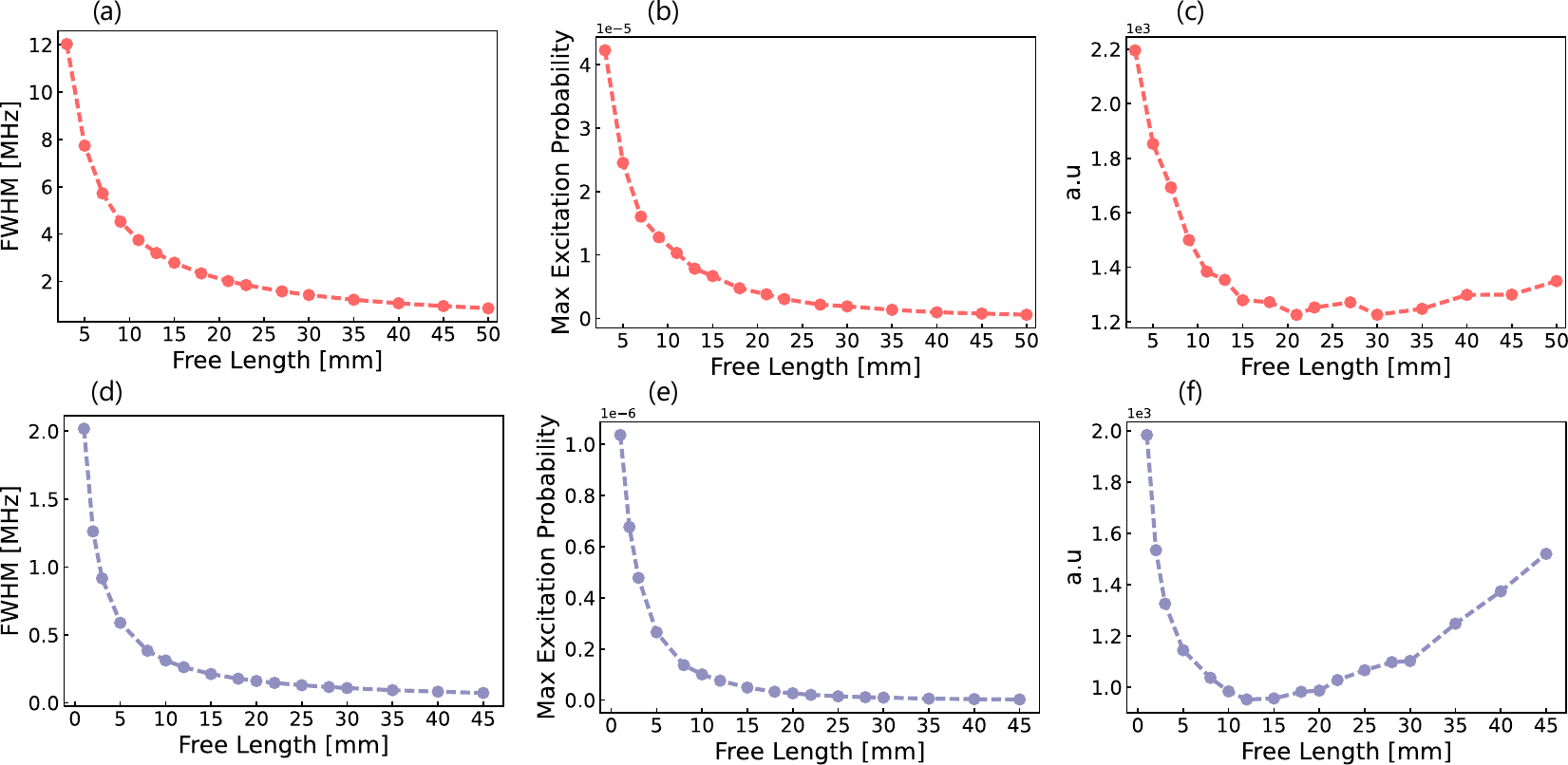}
    \caption{(a) Dependence of the FWHM extracted from the central peak on the free evolution distance between the Ramsey interaction zones for positronium (Ps, red). (b) Maximum excitation probability extracted from the central peak for Ps as a function of the free evolution distance. (c) FOM for Ps, showing the balance between excitation probability and FWHM. (d) Dependence of the FWHM for muonium (M, blue) on the free evolution distance. (e) Maximum excitation probability for M. (f) FOM for M. The optimal free evolution distance minimises the FOM for both systems, corresponding to a range of 15-35 mm for Ps and 8-20 mm fo M.}
    \label{fig:FoM}
\end{figure*}

\subsection{Initial Distribution} \label {velocity}

The results presented by Heiss et al. \cite{Heiss:2024ofn} on time of flight (TOF) measurements of positronium atoms emitted into vacuum from mesoporous silica \cite{Crivelli2010, Cassidy2010} indicate that the Ps velocity can be modelled by a Maxwell-Boltzmann distribution, consistent with the findings of Deller et al. \cite{Deller2015}
This model is centred on an average temperature of \SI{500}{K} for a positron implantation energy of 6 keV. 
Similarly, we modelled the velocity distribution of muonium atoms using a Maxwell-Boltzmann distribution centred on a temperature of \SI{300}{K} \cite{2012-Mesoporous}. 

Ps and M atoms are emitted with a trajectory following a cosine distribution, $\cos(\theta)$, where $\theta$ is the angle relative to the normal of the emission surface \cite{Cassidy2010,PhysRevA.81.012715}. Atom trajectories are generated one by one from a random Gaussian distribution with a sigma of 1 mm, which is the spot size of the $e^+$ beam at ETH Zurich \cite{cooke2015observation, Heiss:2021qxv} and of $\mu^+$ beam at PSI.

\subsection{Laser Parameters}\label{laser}

In our simulation, the laser used to induce the 1S-2S transition in positronium atoms is a CW laser with a Gaussian beam profile. The laser operates at a wavelength of 486 (244)\,nm, with powers of 450 (25)\,W for Ps and M respectively. The laser beam is enhanced by a Fabry-Perot with a build up factor of $\sim$ 20000 for Ps and around 40 for M. The beam waist at the interaction region is approximately 300 $\mu$m in both cases.

To maximise the interaction with the atoms, the first laser, which drives them into a superposition of the ground state and the excited metastable 2S state, is positioned 2 mm downstream from the Ps (M) source. After the free evolution, the atoms pass through the second interaction region.
\subsection{Detection of the 2S atoms}

The detection of the excited state in our setup is performed by exciting the 2S atoms in the 20P state by a second laser at \SI{736}(365){nm} and subsequent detection via field ionisation in a MCP detector for Ps (M) atom \cite{Heiss:2021qxv}. The use of a position-sensitive MCP will enable the reconstruction of their trajectories and time of flight.
By combining the position data with TOF information from MCP, we can effectively calculate the velocities of individual atoms. This capability is essential for correcting the second-order Doppler shift, which would otherwise introduce a significant systematic error in the measured transition frequencies. 
For our simulations, we assume a typical MCP detector’s efficiency of 50\%, a time resolution of 10 ns for Ps and 100 ns for M (the detector resolution is $\sim 1$ ns so this includes the spread introduced by Ps and M out diffusion times) and a spatial resolution of 100 x 100 \SI{}{\micro m^2}. These parameters introduce a maximal error of 0.01 \% for Ps and 0.02 \% for M in the velocity reconstruction (see Figure \ref{fig:setup}).\\

\section{Simulation Results}\label{sec:sim}
\subsection{Ramsey Fringes} \label{sec: Ramsey Fringes}
In this section, we investigate the influence of the free evolution distance on the observed Ramsey fringes, focusing on the relationship between the spatial arrangement of the lasers and the resulting interference patterns. The optimisation of this parameter is discussed in Section \ref{optimization}.

\begin{figure*}[ht]
    \centering
    \includegraphics[width=\textwidth]{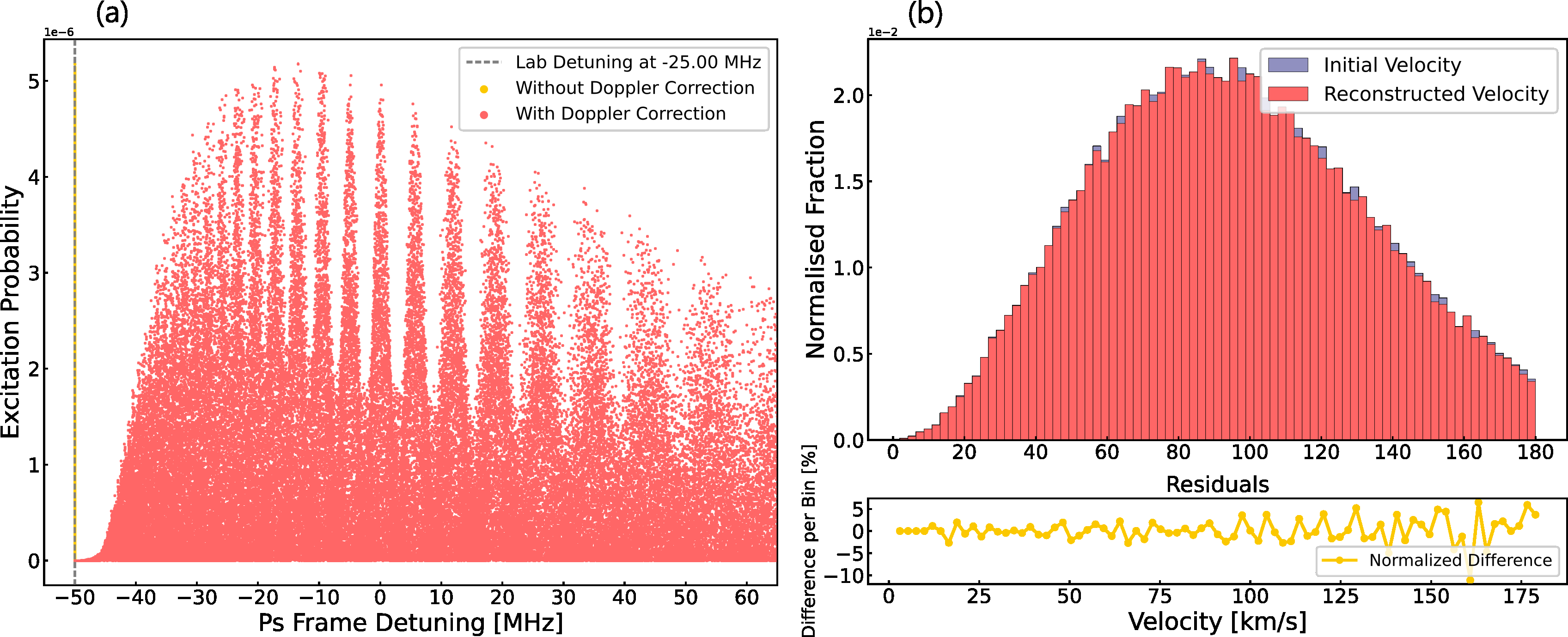}
    \caption{Left: Ramsey-Doppler fringes reconstructed after the second order Doppler shift correction for a laser power of 450 W and a free length of 15 mm. The laser detuning was fixed to -25MHz. Right: Difference between simulated and reconstructed velocity of the Ps atoms. }
    \label{fig:RamseyPsResult}
\end{figure*}

For the ideal case of mono-energetic positronium and muonium trajectories emitted perpendicularly from the SiO$_2$ target, corresponding to temperatures of 500 K for Ps and 300 K for M, the excitation probability for the 2S state as a function of detuning (in MHz) is shown for two different free evolution distances of 2 mm and 5 mm in Figure \ref{fig:Ramsey fringes}. In the Figure, the yellow line represents the single interaction excitation probability for the 2S state, referred to as the ``Rabi" method. At the same time, the red and purple dots correspond to the 2S Ramsey excitation probability for the respective free evolution distances.

The observed detuning shifts of 52 MHz and 500 kHz for Ps and M arise from the combined effects of the second-order Doppler shift and the AC Stark shift. In the atoms reference frame the detuning is given by
\begin{equation}\label{detuning contribution}
    \begin{split}
        \Delta\nu(t) = 2 \nu_L - \nu_{eg} - (\Delta \nu_{ac}(e) 
         \\ -\Delta \nu_{ac}(g)) + \Delta \nu_{2DS},
    \end{split}
\end{equation}
where $ \omega_L$ is the laser detuning, $\Delta \nu_{2DS}$ is the 2nd Doppler, and $\nu_{eg}$ the transition frequency, and $\Delta \nu_{AC}(e)$, $\Delta \nu_{AC}(g)$ for the excited and ground state AC Stark shifts. These shifts and the uncertainty they induce are discussed further in section \ref{sec:syst}.  The plot highlights the distinct behaviour of Rabi and Ramsey methods, with the latter showing improved resolution due to the characteristic interference fringes. A narrowing of the central peak for Ps (M) is visible with the FWHM decreasing with increasing free evolution time from 16.59 (1.26) MHz with 2mm distance to 7.74 (0.59) MHz for 5 mm. A reduction of the excitation probability is observed as the free evolution distance increases due to the decay of the atoms. Consequently, it is essential to identify an optimal free evolution length that minimises the FWHM while retaining a significant excitation probability.

\subsection{Optimization of the Free Evolution Length}\label{optimization}

To estimate the optimal ``free length'' value, we define the figure of merit as:
\[ \text{FOM} = \frac{\text{line width}}{\sqrt{\text{max probability}}} \]
The FOM optimises the balance between line width and event statistics. The local minimum of the FOM indicates the best value for the free length. The plateau corresponding to a local minimum between 15 (10) mm and 35 (20) mm represents the trade-off between narrow line width and maximum probability. At larger distances, the losses from self-annihilation for Ps and from the muon decay for M start to dominate. We select a free length of 15 (10) mm for the Ps (M) simulation, which allows us to achieve a linewidth on the order of 2.5 (0.24) MHz.  

\begin{figure*}[htp]
\includegraphics[width=\textwidth]{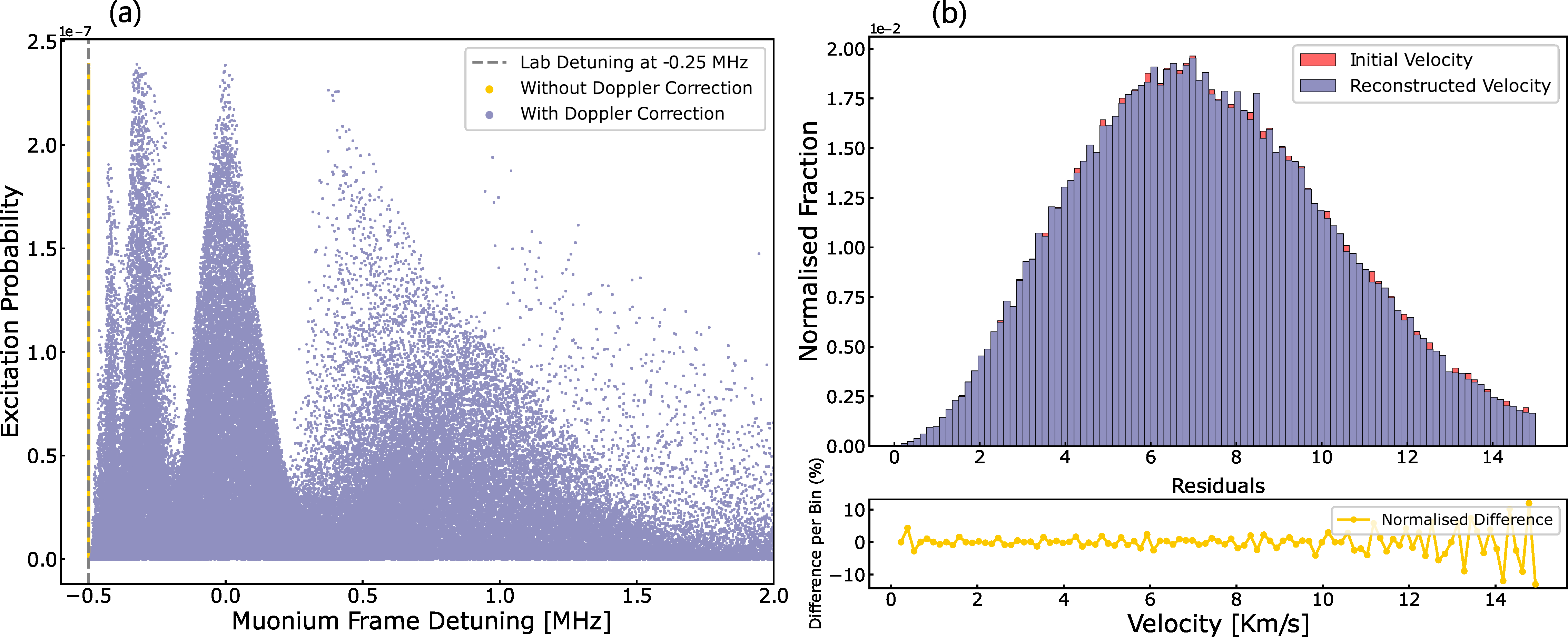}
\centering
\caption{Left: Ramsey-Doppler fringes reconstructed after the second order Doppler shift correction for a laser power of 25 W and a free length of 10 mm. The laser detuning was fixed to -250 kHz. Right: Difference between simulated and reconstructed velocity of the muonium atoms.}
\label{fig:RamseyMResults}
\end{figure*}

\subsection{Second-Order Doppler effect correction}\label{sec: Doppler correction}

In the previous section, we demonstrated that with an MCP, which is sensitive to both the position and the time of intersection of atoms, it is possible to reconstruct the velocity distribution of individual atoms. The fringes are washed out for a thermal distribution with the expected $\cos \theta$ emission angle \cite{Crivelli2010, Cassidy2010,2012-Mesoporous,2016_Khaw_Confinement} from the target. However, by measuring the velocity of the Ps (M) on a per atom basis and correcting for each atom the second-order Doppler shift they experienced, the fringes can be recovered as shown in Figs. \ref{fig:RamseyPsResult}.b and on \ref{fig:RamseyMResults}.b. Interestingly, with this method, the laser frequency does not need to be scanned because the atoms are detuned by the second-order Doppler shift. 

Figures \ref{fig:RamseyPsResult}.a  (\ref{fig:RamseyMResults}.a) show the excitation probability of 100,000 atom trajectories as a function of detuning in the Ps (M) reference frame for atoms with a 500 (300) K distribution. The laser detuning was set to -25.0 (-0.25) MHz from the atom's reference frame. The efficiency of reaching the detector and having interacted with the two lasers is 0.2\% for Ps and 0.6\% for M. The yellow dots represent the probability distribution of atoms without correcting for the second-order Doppler effect. When applying the correction on atom-by-atom base and plotting the excitation probability as a function of the actual detuning experienced by the atoms the Ramsey fringes are clearly visible.

\subsection{Identification of the central peak} 

\begin{figure*}[ht]
    \centering
    \includegraphics[width=\linewidth]{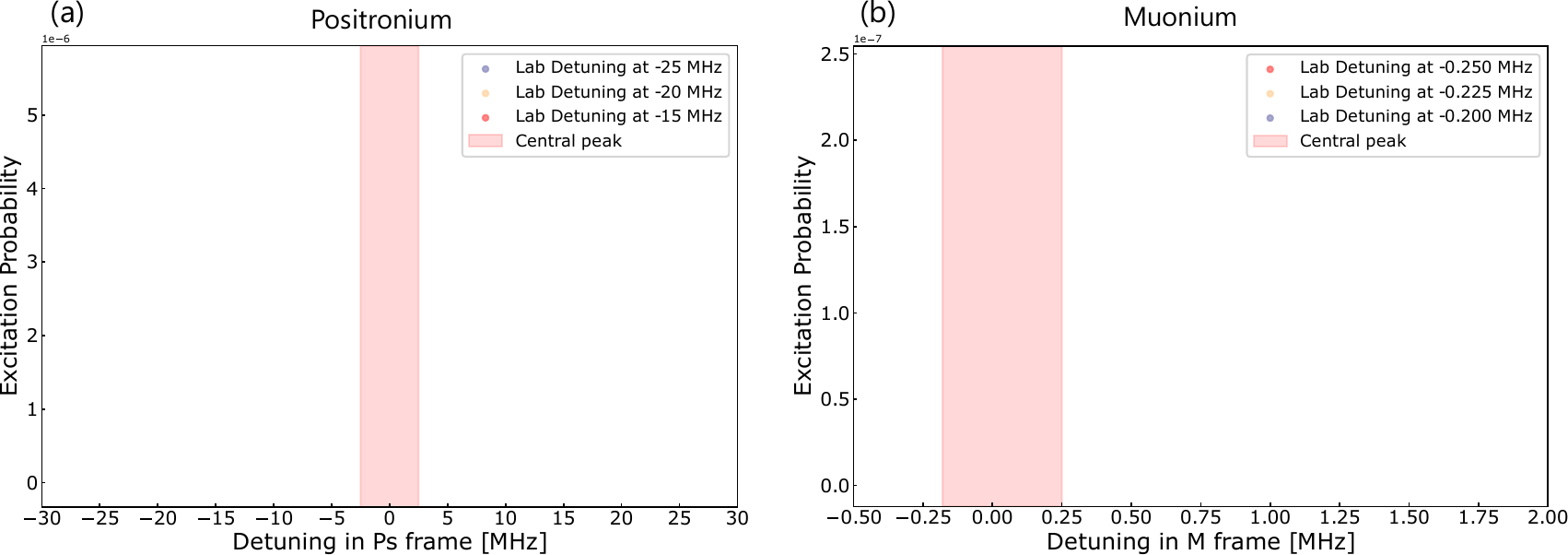}
    \caption{The excitation probability of 100,000 simulated Ps (M) trajectories per detuning plotted against the detuning in the Ps (M) reference frame. The probabilities of Ramsey fringes, with the central fringe at \SI{0}{MHz} highlighted in red - are confirmed by the overlap of Ramsey fringes resulting from three different laser detuning.}
    \label{fig:CentralPeakID}
\end{figure*}

As shown in the previous section, once the Doppler effect is accounted for, the simulated excitation probability reveals the characteristic Ramsey fringes. Interestingly, the central fringe does not have the highest excitation probability, as the red-shifted detunings correspond to slower-moving Ps atoms that spend more time in the laser. To experimentally identify the central fringe, one can measure with multiple laser detunings (in the laboratory frame). When superimposed, as in Figure \ref{fig:CentralPeakID} the central fringe overlaps and adds constructively, while other fringes are shifted relative to the central fringe. We observe a distinct peak at 0 MHz, highlighted in red.

\begin{figure*}[h!]
    \centering
    \includegraphics[width=\linewidth]{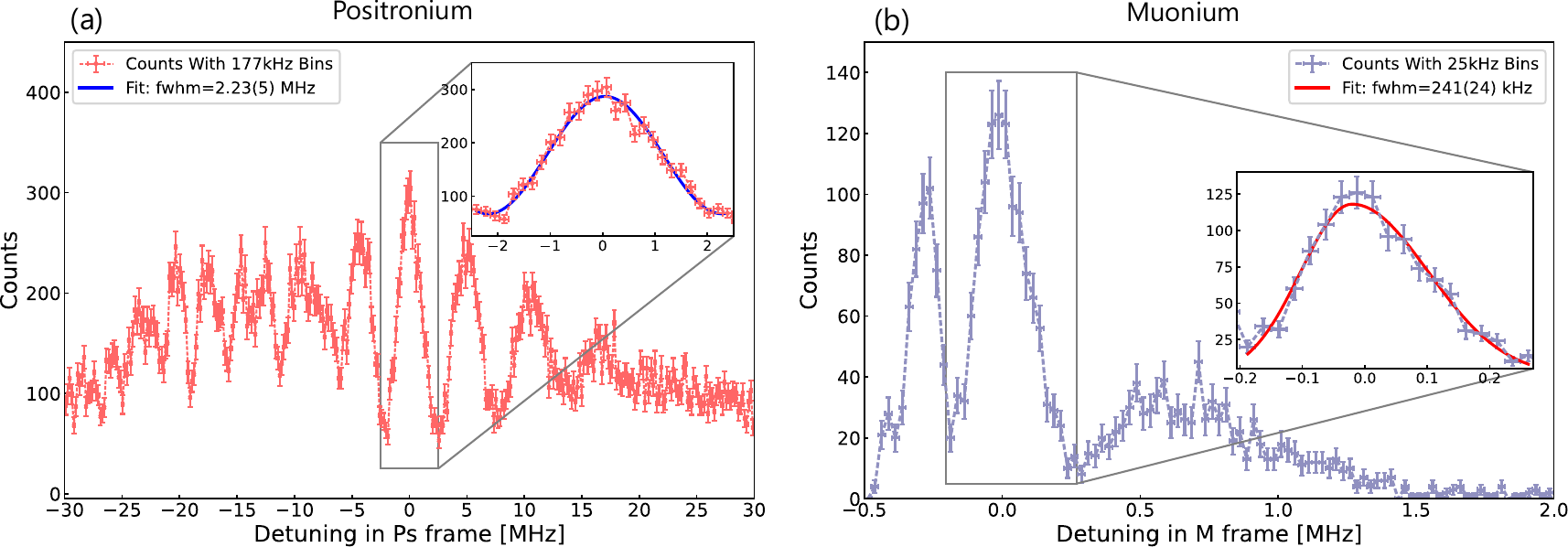}
    \caption{(a) Count rate versus detuning for positronium and muonium (b) after 10 days of measurement (see text for the details on parameters assumptions). The main panels show the detuning dependence of the Ramsey fringes in the respective atomic frames. The insets provide close-ups of the central peaks, fitted to extract the central peak frequency and FWHM: 2.23 (5) MHz for Ps and 241 (24) kHz for M.}
    \label{fig:counts}
\end{figure*}

Figures \ref{fig:counts}.a and \ref{fig:counts}.b show that, for both positronium (left panel) and muonium (right panel), the central peak is distinctly identifiable, as other peaks are washed out due to the detuning shifts when overlaying simulation results with tree laser detunings for Ps (-15MHZ, - 20MHz and -25MHz) and two detunings for M (-200kHz and -250kHz). This constructive interference at the central fringe highlights a clear peak around 0 MHz. A Monte Carlo method is used for detection, where the simulated excitation probability of each atom is compared with a randomly generated probability to decide if the atom is detected. 
These measurements represent simulated data corresponding to a 10-day experimental period. To estimate the resonance frequency of the 1S-2S transition, the central peak is fitted using a function of the form \( A\cos^2(T(x - x_0)) + C \) for positronium. This fit gives a central frequency of \(79 \, \text{kHz}\) with an uncertainty of \(23 \text{kHz}\) and a reduced chi-square value of \( \chi^2_\nu = 1.29 \) with 4 degrees of freedom. For muonium, an asymmetric Gaussian fit is applied, resulting in a centre frequency of \(-20 \text{kHz}\) with an uncertainty of \(9 \text{kHz}\) and a reduced chi-square of \( \chi^2_\nu = 1.08 \) with 4 degrees of freedom. The insets in Figure \ref{fig:counts} show a zoomed-in view of the central peaks, with the fit highlighted.

\subsection{Systematic Uncertainties}\label{sec:syst}
In this section, we describe the estimation of the main systematic uncertainty contributions to the proposed experiments. The results are summarised in Table \ref{tab:uncertainty_budget_M_Ps}. 

Though the proposed scheme accounts for the second-order Doppler shift, uncertainties on the atoms' reconstructed velocities result in a residual Doppler shift.
Using a sample of $10^5$ atoms we calculate the average reconstruction error of the velocity and determine the frequency shift associated with the error in correcting the 2nd order Doppler effect. 
Using all reconstructed detunings, one gets a shift of \SI[separate-uncertainty]{-0.4\pm47.3}{kHz}. However, for the determination of the transition frequency, only the central fringe contributes to the fit. As such, we only need to consider atoms whose total detuning (see \eqref{detuning contribution}) is in the central fringe. We consider the reconstructed detunings of $\pm\SI{5}{MHz}$ central. In this range, the uncertainty from the residual second Doppler is $\SI[separate-uncertainty]{0.1\pm17.0}{kHz}$. The residual Doppler is relatively insensitive to the choice of frequencies for the central fringe. Even when allowing detunings leading to the inclusion of an additional fringe (around $\pm\SI{10}{MHz}$) the uncertainty due to the Doppler is still only about $\SI{17.4}{kHz}$. 

\begin{figure}[h]
    \centering
    \includegraphics[width=0.9\linewidth]{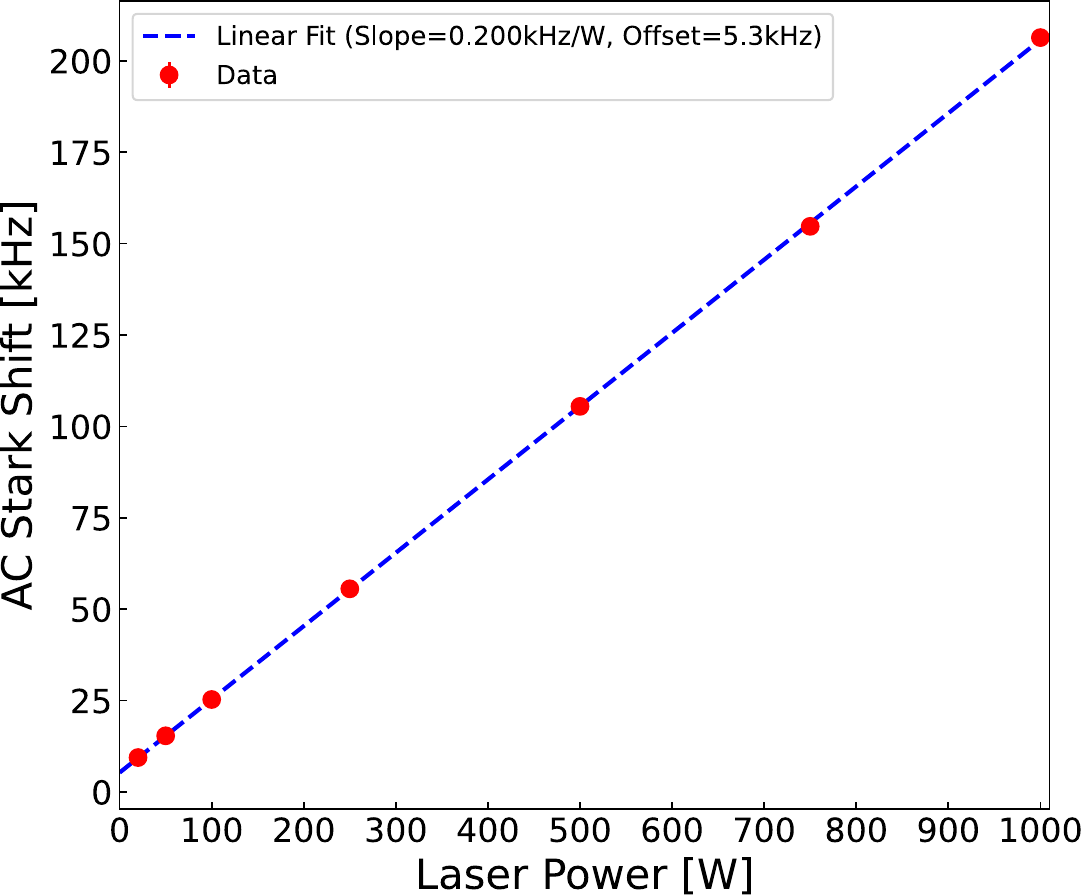}
    \caption{AC Stark shift extrapolation of Ps.}
    \label{fig:Ps_AC_Stark}
\end{figure}

The frequency shift due to the AC Stark shift induced by the exciting laser field is $\Delta \nu_{AC}(e)= \beta_{AC}(e) \cdot S\cdot  I(t)$ for the ground state and likewise for the excited one $\Delta \nu_{AC}(g)$. The coefficients for hydrogen are $\beta_{AC}(e)=-2.67827\times 10^{-5}$ Hz/(W/m$^2$)$^{-1}$ and $\beta_{AC}(g)=1.39927\times10^{-4}$ Hz/(W/m$^2$)$^{-1}$, which are rescaled for Ps and M by multiplying with a factor, $S=(m_e/\mu)^3$, with $\mu$ the reduced mass of the atom \cite{Haas2006TwophotonED}. In Ramsey spectroscopy the AC Stark shift is expected to be suppressed by a factor $\omega_0/D$, where $D$ is the free evolution distance and $\omega_0$ the beam waist of the laser. Intuitively, the suppression originates from the fact that the atoms spend significantly more time acquiring their phase in the field-free region, where there is no Stark shift, than in the laser. The effect has been described in \cite{borde1983density_ac_suppr} and corroborated by the MC simulation. Taking Ps for example, the expected Stark shift for a \SI{450}{W} laser with a \SI[separate-uncertainty]{310}{\micro m} waist and taking a conservative uncertainty of 10\% on the laser power one gets an uncertainty of about \SI{400}{kHz} on the AC Stark shift. However, in the Ramsey scheme, the inclusion of the factor $\omega_0/D$ for a \SI[separate-uncertainty]{15\pm 1}{mm} free propagation region results in  only \SI{9.9}{kHz} uncertainty on the shift, which is negligible compared to aimed precision of the experiment. The estimate assumes that the Ps experiences the maximal intensity, i.e. always traverses the centre of the Gaussian beams. The results from a simulation of monodirectional, monoenergetic Ps are shown in Figure \ref{fig:Ps_AC_Stark} where the AC Stark shift is determined by simulating at several powers. Using the extracted slope and assuming a power of \SI[separate-uncertainty]{450\pm45}{W}, the uncertainty on the shift is calculated to be \SI{9}{kHz}, which is similar to the value calculated using the $\omega_0/D$ suppression factor. For M, we use \SI{25}{W} (with 10\% uncertainty), \SI[separate-uncertainty]{300}{\micro m} and \SI[separate-uncertainty]{8\pm 1}{mm} free propagation distance. 

Two additional systematic uncertainties worth mentioning are the DC stark shift and the Zeeman shift, both of which we estimate to have negligible contributions. The 1S-2S transition is sensitive only to the quadratic DC Stark shift, which in atomic units is given by $\Delta \nu_{DC} =- \alpha_p/4\pi \mathcal{E}^2$, where $\mathcal{E}$ represents the stray electric fields in the excitation chamber, $\alpha_p = 576 a_0^3$ the polarisability of hydrogen in the 2S state with $a_0$ the Bohr radius. Assuming stray fields of about \SI{0.2}{V/cm}, the resulting shift in hydrogen and muonium would be about \SI{0.14}{kHz} \cite{sandberg1993research}. For Ps the polarisability is 8 times larger due to the 2 times larger Bohr radius and therefore the shift is also negligible. Likewise, the second order Zeeman shift $\Delta \nu_{B} =-\frac{\alpha^2 \mu_B }{4h}B$
where $\mu_B$ is the Bohr magneton and $\alpha$ the fine structure constant \cite{sandberg1993research}. For the Earth magnetic field $B\sim 50\mu$T, this results in a frequency shift which is well below \SI{1}{Hz}. 

Using a frequency comb referenced to a satellite with an Allan deviation of the order $10^{-12}$, one can reach a precision on the laser frequency of about 1 kHz. It is worth noting that using the fiber link from METAS \cite{Husmann:2021juv}, to which ETH is already connected, this could be reduced by orders of magnitude. 

\begin{table}[h!]
\centering
\caption{Summary of systematic uncertainties
}
\begin{tabular}{|l|c|c|l|}
    \hline
    \textbf{Uncertainty Source} & \textbf{Ps} [kHz] & \textbf{M} [kHz]\\
    \hline
    Residual 2nd Doppler & 17.4 & 1.4 \\
    AC Stark Shift & 9.9 & 0.2  \\
    DC Stark Shift & 1.6 & 0.14 \\
    Zeeman & <0.001&  <0.001\\
    Laser Frequency& 1 & 1 \\
    \hline
    Sum & 20 & 1.7 \\
    \hline
\end{tabular}
    \label{tab:uncertainty_budget_M_Ps}
\end{table}

\section{1S-2S Ramsey Spectroscopy with a SFHe Muonium Source} \label{SFHe}
In this section, we discuss the possibility of utilising a superfluid helium (SFHe) muonium source to perform Ramsey spectroscopy. The experimental scheme proposed for the SFHe formation target differs from the SiO$_2$ approach, as the M atoms are ejected with an almost delta function-like longitudinal velocity distribution peaked at 2175 m/s \cite{Soter2024}. The velocity spread is predicted to fall below the Landau critical velocity of Muonium in SFHe, which we take conservatively as $\sim100$ m/s, following the estimations from \cite{Soter_2021_SFHe, osti_6584671}. The fact that the velocity of Muonium is well known releases the measurement scheme from the need of reconstructing the atom velocities. Therefore, once the atoms are excited, the detection of the 2S states can follow the method currently used in the Mu-MASS experiment \cite{cortinovis2023update}. Here, the atom is ionised by a pulsed laser (\SI{355}{nm}), and the ionised muon is collected on an MCP in a separate detection area, shielded from the laser. Due to the low excitation rates, this experiment will benefit from the higher muon rates that the upgraded beamlines at PSI will deliver. The predicted muonium rate per second using a SFHe target of 10x1 mm\textsuperscript{2} are shown in the first two rows of Table \ref{tab:PredMRates}. The background rate in the current 1S-2S experiment was suppressed and measured to be zero in 65 hours of collected data \cite{IreneThesis}, however in the proposed measuring scheme, potential new background sources related to the SFHe and the new spatial configuration of the UV laser may need to be addressed.  

\begin{figure}[ht]
    \centering
    \includegraphics[width=\linewidth]{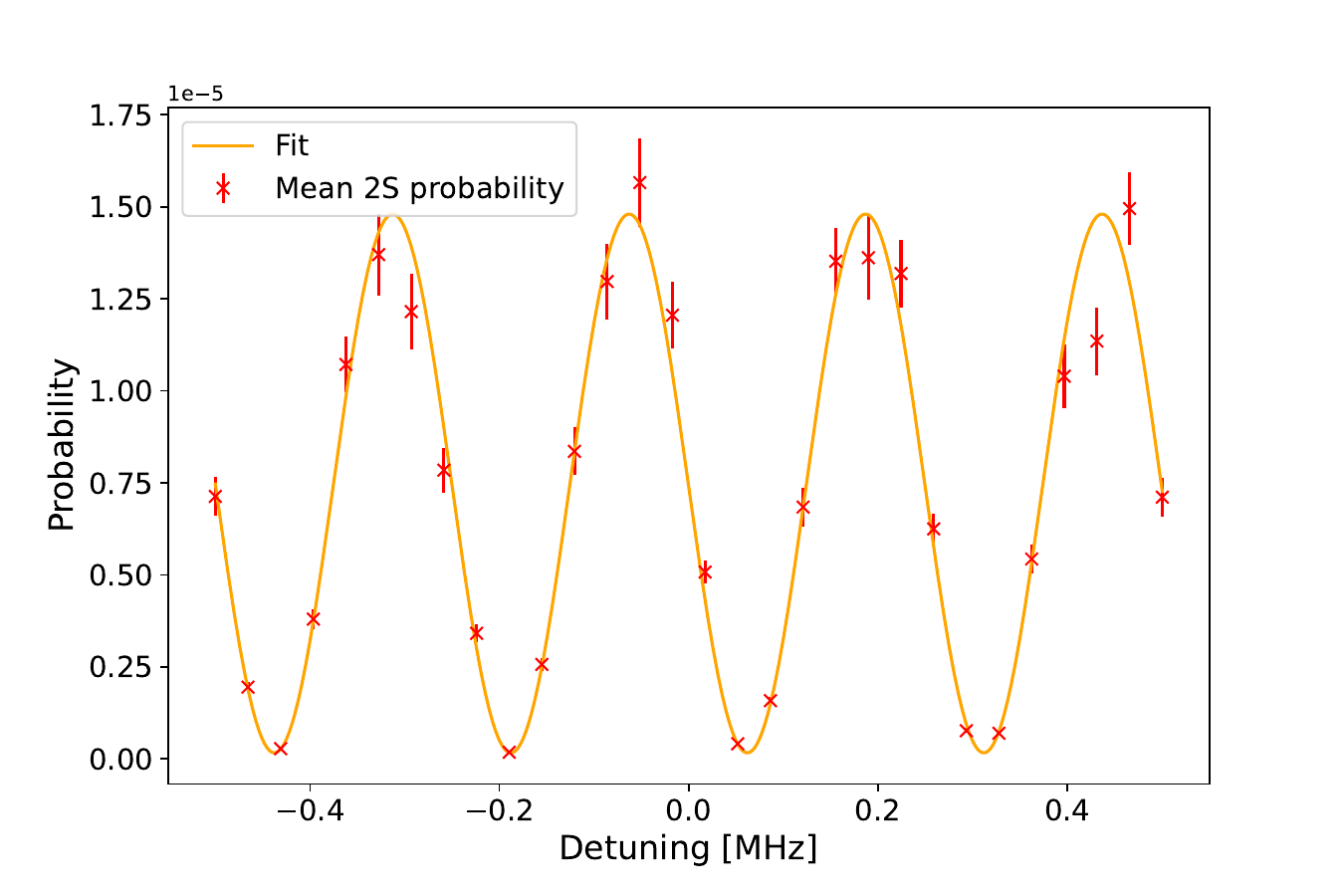}
    \caption{Simulated Ramsey fringes obtained for the optimised scenario with a free length of 8 mm. The 244 nm laser power is 25 W and the waist is 0.35 mm. The $\mu^+$ to muonium converter is a 10 mm x 1 mm SFHe target. Adapted from \cite{IreneThesis}}
    \label{fig:SFHeRamseyFringes}
\end{figure}
A simulation with a SFHe target was conducted using \SI{25}{W} of power circulating in the cavity and a \SI{350}{nm} waist, as these are the best parameters at which the current M 1S-2S experiment could run in a stable operation mode \cite{Zhadnov:23}. Higher powers up to \SI{40}{W} were achieved with this setup, though only for a limited time, which suggests potential for further improvements with additional development efforts. The target was modelled as a $\SI{10}{mm} \times \SI{1}{mm}$ rectangle. In Figure \ref{fig:SFHeRamseyFringes} a simulation of the Ramsey parameters is seen for an optimised free propagation length of \SI{8}{mm}. In this simulation, the first laser interaction zone is placed \SI{2}{mm} from the SFHe target to optimise signal rate. This experiment results in typical Ramsey fringes, which can be seen in Figure \ref{fig:SFHeRamseyFringes}. The simulation was done accounting for 24 hours at the upgraded HIMB-3 beamline with about $10^5 \, \SI{}{\micro^+/s}$ for 100 points at each of the 30 frequency points scanned. The relatively low statistics despite the high rate comes from a combination of a low formation fraction ($\sim$10\% \cite{Soter2024}) and excitation probability. 
When performing the experiment, it might be hard to identify the central peak as the fringes are very similar in height. The central peak could be identified experimentally by changing the free propagation distance and overlaying the datasets on each other. The free propagation distance could be changed by moving the position of the cavity mirrors of the second interaction region by using a piezoelectric stage. As only the position of the central fringe remains unaffected by the free propagation distance, this helps in identifying it. Measuring at the optimal free propagation length and by taking into account the ionisation and detection efficiencies to calculate the final event rates, 24h would be needed to collect the statistics reported in Figure \ref{fig:SFHeRamseyFringes}, resulting in a FWHM of the central fringe of about \SI{120}{kHz}.

Assuming ten days of measuring beamtime at PSI, the linecenter can be then determined with a precision of about 1kHz. 
This projection relies on the implementation of the SFHe target into the setup and the realization of the slow muon rate upgrades. If successful, this method would provide a concrete way forward to use the proposed method for precision studies of muonium.

\begin{table}
    \centering
    \begin{tabular}{|l|c|l|}
        \hline
        Scenario & $\text{R}_\text{M}$ [\SI{}{s^{-1}}] \\
        \hline
         HIMB-5, SFHe &  $\SI{4e4}{}$ \\
         HIMB-3, SFHe &  $\SI{1e5}{}$ \\
         HIMB-3 + MuCool, SiO${}_2$ &  $\SI{3e4}{}$ \\
         \hline
    \end{tabular}
    \caption{Predicted M rates for planned upgrades at PSI - both regarding the muon beams and the M formation method - for a target size of 10x1 mm\textsuperscript{2}. Adapted from \cite{IreneThesis}.}
    \label{tab:PredMRates}
\end{table}

\section{Conclusion}\label{conclusion}

With a realistic simulation, we have demonstrated the feasibility of Ramsey spectroscopy combined with correction of the second-order Doppler shift for the 1S-2S transition in Positronium and Muonium and propose an experimental scheme to achieve this measurement. 
The Monte Carlo simulation includes modelling of positronium and muonium formation, laser interactions, and detection processes. The optical Bloch equations are numerically solved for both excitation regions and free evolution, and incorporate effects such as photoionisation, the AC Stark effect, and atomic decays. Furthermore, we propose a novel approach to obtain Ramsey fringes with only a single laser detuning, by utilising the velocity distribution of atoms to scan the line. 

Our results show that two-photon Ramsey-Doppler spectroscopy has the potential to enhance the precision of positronium spectroscopy by two orders of magnitude over the 1993 measurements by M. Fee et al. \cite{Fee1993} and four orders of magnitude over V. Meyer et al. Measurement of the 1S-2S energy interval in muonium \cite{2000_Meyer}. Additionally, our simulation extends to testing standard Ramsey spectroscopy on muonium with a superfluid helium source, indicating that a kHz-level determination might be achievable.

Combining Ramsey-Doppler spectroscopy with one-dimensional laser cooling \cite{Shu2024,PhysRevLett.132.083402} to collimate the positronium beam would increase significantly the available statistics providing even higher precision for such measurements. While 3D laser cooling for positronium may become feasible in the near future, it remains highly challenging for muonium due to the much shorter wavelength. This makes the Ramsey-Doppler technique particularly well-suited for advancing spectroscopy of the muonium system.

\section*{Acknowledgments}\label{sec6}
We are in debt to Thomas Udem for motivating us to explore the potential of optical two-photon Ramsey spectroscopy. This work is supported by the Swiss National Science Foundation under grant 219485. 

\section*{Authors contribution}
All authors contributed to the study. Material preparation, data collection and analysis were performed by all authors. All authors read and approved the final manuscript.

\section*{Data availability statement}
This manuscript has associated data in a data repository. [The Authors datasets generated during and/or analysed during the current study are available from the corresponding author on reasonable request].

\bibliographystyle{unsrt}
\bibliography{sn-article}

\end{document}